\documentclass[conference]{IEEEtran}
\IEEEoverridecommandlockouts
\usepackage{cite}
\usepackage{amsmath,amssymb,amsfonts}
\usepackage{algorithm}
\usepackage{algpseudocode}
\usepackage{tikz}
\usepackage{quantikz}

\usepackage{graphicx}
\usepackage{textcomp}
\usepackage{comment}

\usepackage{booktabs}

\usepackage{soul}

\usepackage{array}
\usepackage{xcolor}
\def\BibTeX{{\rm B\kern-.05em{\sc i\kern-.025em b}\kern-.08em
    T\kern-.1667em\lower.7ex\hbox{E}\kern-.125emX}}

\definecolor{myblue}{RGB}{0, 170, 200}
\definecolor{mygreen}{RGB}{30, 200, 70}

\begin{document}

\title{Distributed Quantum-Enhanced Optimization: A Topographical Preconditioning Approach for High-Dimensional Search\\
%%\thanks{Identify applicable funding agency here. If none, delete this.}
}

\author{
\IEEEauthorblockN{
Dominik So\'os\IEEEauthorrefmark{1}, 
Marc Paterno\IEEEauthorrefmark{1}\IEEEauthorrefmark{2},
John Stenger\IEEEauthorrefmark{1}\IEEEauthorrefmark{3}, 
%Daniel Gunlycke\IEEEauthorrefmark{3}, 
Nikos Chrisochoides\IEEEauthorrefmark{1}\IEEEauthorrefmark{4}
}

\IEEEauthorblockA{
\IEEEauthorrefmark{1}Department of Computer Science, Old Dominion University, Norfolk, VA, USA \\
\IEEEauthorrefmark{2} Computational Science and Artificial Intelligence Directorate,  Fermi National Accelerator Laboratory, Batavia, Illinois, USA \\ 
\IEEEauthorrefmark{3}Chemistry Division, Naval Research Laboratory, Washington, D.C., USA \\
\IEEEauthorrefmark{4}Department of Physics, Old Dominion University, Norfolk, VA, USA
}
}

\maketitle

\begin{abstract}
Optimization problems become fundamentally challenging as the number of variables increases.
Because the volume of the search space grows exponentially, classical algorithms frequently fail to locate the global minimum of non-convex functions.
While quantum optimization offers a potential alternative, mapping continuous problems onto near-term quantum hardware introduces severe scaling limits and barren plateaus. 
To bridge this gap, we propose the Distributed Quantum-Enhanced Optimization (D-QEO) framework. 
Instead of forcing the quantum processor to find the exact minimum, we use it simply as a topographical preconditioner.
The QPU maps the landscape to locate the most promising basin of attraction, generating high-quality seed points for a classical GPU-accelerated solver to refine.
To make this approach viable for utility-scale problems, we exploit the mathematical structure of separable functions.
This allows us to cut a 50-qubit (i.e., $2^{50}$) global search space into independent and manageable sub-spaces using 5-qubit subcircuits.
By executing these fragments concurrently with CUDA-Q, we completely bypass the overhead of cross-register entanglement and classical tensor knitting for separable functions. 
Benchmarks on the 10-dimensional Rastrigin and Ackley functions show that D-QEO prevents the exponential failure rates observed in purely classical algorithms. 
Furthermore, this quantum warm-start significantly reduces the number of classical BFGS iterations required to converge, providing a highly practical blueprint for utilizing near-term quantum resources in complex global search.
%
% MFP: I suggest we leave the following line commented out until we have the algorithm in a public repository.
%
% Our code is publicly available at X.
\end{abstract}


\section{Introduction}
Global mathematical optimization is a vital component of modern scientific and industrial applications, ranging from machine learning and financial modeling to the simulation of complex physical systems.
Despite significant progress in classical algorithmic design, the ``curse of dimensionality'' continues to impose a fundamental limit on the scalability of global search strategies.
As the number of dimensions increases, the volume of the search space grows exponentially, making the number of sample points required to guarantee convergence to the global minimum prohibitively large.

Recent advancements in classical optimization, such as the {\sc Zeus} framework~\cite{soos2025zeus}, have sought to mitigate these challenges by leveraging the massive parallelism of modern Graphical Processing Units (GPUs).
By combining stochastic Particle Swarm Optimization (PSO) for global exploration~\cite{kennedy1995particle} with the quasi-Newton Broyden-Fletcher-Goldfarb-Shanno (BFGS) method for local refinement~\cite{broyden1970convergence,fletcher1970new,goldfarb1970family,shanno1970conditioning}  and utilizing forward-mode Automatic Differentiation (AD) for accurate gradient calculation~\cite{rall1996introduction}, ZEUS provides a high-throughput pipeline for non-convex landscapes.
However, even with such high-performance classical methods, the exponential scaling of the search space remains an intractable barrier (see Figure~\ref{fig:motivation}).
Empirical studies on the multimodal Rastrigin function~\cite{rastrigin1974systems} demonstrate that as dimensionality grows, the probability of successful convergence to the global minimum basin decays exponentially.
This failure mode, where the number of correct solutions reaches effectively zero by just 10 dimensions, suggests that purely classical approaches are fundamentally limited even for moderately dimensional spaces.

In real-world applications, one often does not know the number of local minima present in the function being minimized.
For such problems it is often necessary to try many different starting points for the minimization in order to achieve sufficient confidence that one has obtained the correct global minimum.
One such example in high-energy physics is the recent neutrino mixing analyses in the NOvA experiment~\cite{bian2013nova,Acero:2022nqx}.
This analysis relies on the profiled Feldman-Cousins (FC) technique~\cite{NOvA:2022wnj}.
In the NOvA analysis, calculating confidence intervals at the 3-sigma level required more than 20 million core-hours to generate a single two-dimensional contour plot~\cite{NOvA2022Poster}.

As future high-precision experiments like DUNE~\cite{abi2020volume} target 4-sigma confidence intervals, the computational effort of generating, fitting, and searching millions of pseudoexperiments across the parameter space is projected to increase by a factor of more than 40. 
Currently, tackling this exascale-level challenge requires developing parallel GPU-accelerated algorithms tailored for next-generation supercomputers.
By utilizing the Quantum Processing Unit (QPU) as a topographical preconditioner to map global basins and warm-start classical GPU swarms, our hybrid approach directly addresses the structural inefficiencies of these massive parameter searches, paving the way to make 4-sigma analyses and other high-demand scientific applications computationally feasible down the road.

The emergence of hybrid quantum-classical algorithms offers a promising alternative to bypass these classical bottlenecks. %~\cite{ornl, nvidia, ibm-japan}. 
Quantum computing introduces a Hilbert space that can inherently represent exponentially large states, providing a platform for global operations on the entire optimization landscape.
However, naive translations of classical optimization to the quantum regime often struggle with the ``barren plateau'' problem, where gradients vanish exponentially in high-dimensional Hilbert spaces, and the high sampling costs associated with unstructured search.
 To achieve a true quantum advantage, it is necessary to move beyond unstructured exploration and adopt a mapping strategy that reflects the problem's dimensionality while leveraging the mathematical structure of the objective function \cite{haidar2025non}.

In this paper, we propose a ``quantum-aware'' framework, Distributed Quantum-Enhanced Optimization (D-QEO), which achieves scalability by utilizing a register-based dimensional mapping rather than a particle-based mapping. 
By assigning each of the 10 dimensions of the Rastrigin function to its own dedicated 5-qubit register, the framework utilizes a total of 50 qubits. 
This collective tensor-product space allows us to represent the entire $2^{50}$ discretized search space in a single quantum wave-function probe. 
%The primary novelty of our approach is an architectural role reversal, where we deploy the QPU strictly as a coarse-grained ``probability-mapmaker'' to identify basins of attraction, leaving high-resolution refinement to classical GPU engines. 
The primary novelty of our approach lies in formulating the hybrid quantum-classical workflow as a 2-level data- and circuit-decomposition~\cite{maciejunes2025solvinglargescalevehiclerouting, billias2025utilityscalequantumedgedetection} as a preconditioner followed by a classical HPC search~\cite{soos2025zeus}. Rather than executing a monolithic search, we employ a search-and-prune technique: the QPU is deployed strictly to execute a coarse-grained global search that prunes subspaces and isolates promising basins of attraction. This topographical decomposition drastically reduces the search volume, delegating only the localized, high-resolution refinement to classical GPUs.

% To handle the resulting hardware requirements on near-term infrastructure, we implement a 2-level decomposition involving spatial data partitioning managed by a Dask distributed scheduler and circuit-level wire-cutting to execute 5-qubit subcircuit fragments in parallel.
% The primary contribution of this work is the development of a hybrid optimization protocol that utilizes the QPU as a hardware-level preconditioner to improve the convergence of classical GPU swarms.
We demonstrate that by identifying basins of attraction through a variational quantum probe, we can substantially reduce the required number of classical iterations.
Furthermore, we provide evidence for the conjecture that increasing qubit resolution (i.e., finer data decomposition and thus increasing the importance of cutting/knitting~\cite{tang2025enabling}) directly reduces the energy footprint of global search by replacing dissipative classical iterations with energy-preserving unitary quantum operations \cite{morimoto2024continuous, lubowe2022best}. 
This integration ensures that the convergence rate remains robust even as problem dimensions scale into regimes that have remained inaccessible to traditional (i.e., classical) optimization theory.

Our contributions are as follows.
\begin{itemize}
    \item \textbf{Topographical Preconditioning}: We introduce the use of QPU as a preconditioner, identifying global basins of attraction to ``warm-start'' classical GPU optimizer.
    \item \textbf{Exponential Volume Reduction}: We provide evidence that our framework reduces the search volume exponentially (see Table~\ref{tab:volume_reduction}).
    \item \textbf{One-level decomposition (dimension-wise)}: We partition the problem space into independent quantum subregisters (for separable functions) as opposed to particle (data) decomposition.
    \item We show that our implementation outperforms the state-of-the-art HPC classical algorithm~\cite{soos2025zeus} in terms of the number of correct solution for both the  Rastrigin and Ackley functions and provide evidence that this approach is promising for non-separable functions such as the Himmelblau function. 
\end{itemize}

\section{Background}
\subsection{Curse of Dimensionality}
The primary challenge in global mathematical optimization remains the ``curse of dimensionality'', where the volume of the search space grows exponentially with each added dimension.
In non-convex landscapes with multiple local minima, classical search algorithms must navigate a solution space that becomes increasingly sparse as the number of dimensions $d$ increases.
This scalability bottleneck is most evident in multimodal test functions like the Rastrigin function, which has $11^d$ local minima within the standard range $[-5.12, 5.12]$.
We consider the global optimization of a non-convex function $f(\mathbf{x})$ where $\mathbf{x} \in \mathbb{R}^d$ with $d$ is the dimensionality. 
A canonical benchmark is the Rastrigin function:
\begin{equation}
f(\mathbf{x}) = A\times d + \sum_{i=1}^{d} \left[ x_i^2 - A \times \cos(2\pi x_i) \right]
\end{equation}
For $A=10$, this function possesses a global minimum at $\mathbf{x}=\mathbf{0}$, surrounded by a dense field of local minima. Assuming each dimension $x_i$ is bounded by an interval of length $L$, the total volume of the search space $\mathcal{V}$ increases exponentially as $L^d$.
Consequently, for a constant particle density $\rho$, the number of required classical samples $N = \rho \mathcal{V}$ grows exponentially. 
Experimental data shows that for $d=10$, classical swarm success rates drop significantly ($d \ge 5$) as particles fail to initialize within the global basin of attraction~\cite{soos2025zeus}.

While modern GPU-accelerated methods like {\sc Zeus} can process thousands of starting points simultaneously using PSO and BFGS methods, they remain fundamentally constrained by the probability of landing in the basin of the global minimum. 
%Experimental data for the Rastrigin function demonstrates that the number of successful convergences---defined as an optimization result with a Euclidean error less than 0.5---degrades drastically as the dimensionality increases. 

\begin{figure}
    \centering
    \includegraphics[width=0.5\textwidth]{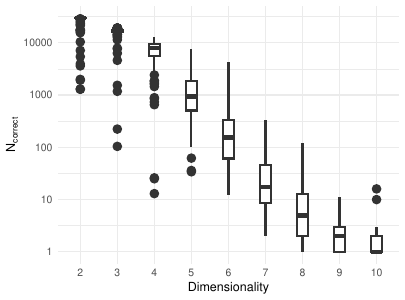}
    \caption{Box and whisker plot showing performance degrades drastically for a classical solver for the Rastrigin function as the dimensionality of the problem increases when using the same number of particles}
    \label{fig:motivation}
\end{figure}

As shown in Figure~\ref{fig:motivation}, when the number of dimensions is increased from 2 to 10 while maintaining a constant number of particles ($10^5$), the count of correct solutions ($N_{correct}$) follows a clear exponential decay. 
This rapid decay highlights that for a 10-dimensional problem (containing approximately 26 billion local minima), the number of particles that successfully locate the true global minimum basin is effectively zero.
This exponential failure rate suggests that simply increasing classical computational resources or particle counts is an insufficient strategy for high-dimensional global search.

\subsection{Non-differentiable Landscapes: The Ackley Function}
To test the resilience of D-QEO, we also evaluate a separable variant of the widely benchmarked Ackley function~\cite{ackley2012connectionist}: 
\begin{equation}
f(\mathbf{x}) = \sum_{i=1}^d \left( -20 \exp\left(-0.2 |x_i|\right) - \exp\left(\cos(2\pi x_i)\right) + 20 + e \right)
\end{equation}

Unlike the standard Ackley function that introduces coupling with a global radial $\ell_2$-norm in the first term and an averaged exponential term, this modified variant is a sum of independent one-dimensional Ackley slices.
This separability allows compatibility with current decoupled\footnote{Future work will address weakly and partially coupled paradigms, outside the scope of this paper.} distributed circuit decomposition methodology for now, allowing $d$ independent quantum circuits to process the landscape without requiring cross-circuit entanglement. 

The separable Ackley function preserves the challenging geometry of its native counterpart. 
It still features a deep central funnel leading to the global minimum, surrounded by a nearly flat outer plateau with many shallow local minima. 
We show that this topography is still difficult for classical exploration (see Figure~\ref{fig:rastrigin_ncorrect} and Figure~\ref{fig:ackley_ncorrect}). 

Furthermore, the absolute value component $|x_i|$ makes the functions non-differentiable at the origin, which is where the global minimum is located. 
This point of non-differentiability restricts purely gradient-based classical approaches when they approach convergence.

\begin{comment}
\subsection{Quantum Mapping and Hamiltonian Derivation}
We discretize each dimension $x_i$ into a register of $K$ qubits. The coordinate mapping is defined as:
\begin{equation}
\hat{x}_i = x_{min} + \Delta \sum_{k=0}^{K-1} 2^k \hat{n}_{i,k}
\end{equation}
The total Hamiltonian $\hat{H}$ is constructed by embedding the objective function $f(\mathbf{x})$ onto the qubit registers. For the Rastrigin function, the Hamiltonian is perfectly separable into $d$ independent clusters, $\hat{H}_{tot} = \sum \hat{H}_i$, which allows for lossless circuit cutting at dimensional boundaries. For the Ackley function, we utilize a Taylor expansion or a lookup-table approach to map the coupled exponential terms, leveraging the function's radial symmetry to maintain efficient circuit widths.
\end{comment}

\section{Literature Review}
A comprehensive review of existing quantum and quantum-inspired algorithms reveals critical gaps that this paper aims to fill. 
We examine the Quantum Approximate Optimization Algorithms (QAOA), Variational Quantum Eigensolvers (VQE), and Quantum Particle Swarm Optimization (QPSO).

\subsection{Quantum Approximate Optimization Algorithm}
QAOA is an iterative method designed primarily for combinatorial optimizations such as Max-Cut \cite{farhi2014quantum,guerreschi2019qaoa,shaydulin2019evaluating}. 
It applies a sequence of cost and mixing Hamiltonians to find the ground state of the problem.
While theoretically sound for discrete structures, NISQ-era implementations of QAOA have shown only marginal improvements over classical baselines and are highly sensitive to parameter settings and the structural noise of the hardware \cite{shaydulin2019evaluating,mcclean2018barren,cerezo2021effect}.
Furthermore, QAOA is difficult to adapt to continuous high-dimensional optimization without significant discretization overhead that often leads to deep, decoherence-prone circuits \cite{morimoto2024continuous,verdon2019quantum,luna2025implementation}.
D-QEO addresses this gap by utilizing a register-based dimensional mapping. 

\subsection{Variational Quantum Eigensolver}
VQE is a prominent primitive for near-term quantum advantage, particularly in quantum chemistry and molecular ground-state approximations \cite{peruzzo2014variational,tilly2022variational,kandala2017hardware}. 
It prepares a parameterized quantum state and measures the expectation value of a Hamiltonian, using a classical optimizer to tune the parameters. 
Although VQE displays resilience against certain noise types, its application to continuous mathematical function optimization remains largely under-explored compared to its use in electronic structure calculations~\cite{tilly2022variational,intoccia2025quantum,haidar2025non}.
Furthermore, existing studies rarely scale beyond 20-qubits~\cite{tilly2022variational,kandala2017hardware,gunlycke2024cascaded}.
D-QEO fills this gap by leveraging the mathematical separability of high-dimensional functions to enable a 50-qubit implementation through scalable circuit cutting~\cite{bravyi2016trading,peng2020simulating}.

\subsection{Quantum-Inspired and Hybrid PSOs}
Quantum-behaved PSO (QPSO) and other quantum-inspired variants simulate quantum dynamics on classical hardware to improve global search \cite{sun2004particle,fang2010review,sun2012quantum}.
While these methods improve upon standard PSO \cite{kennedy1995particle} by introducing mechanisms such as probabilistic tunneling and improved population diversity mechanisms \cite{mikki2006quantum,coelho2010gaussian,fan2026particle}.
However, as they do not utilize true quantum superposition or entanglement, they remain fundamentally bound by classical computational scaling limits \cite{li2011hybrid,liang2005dynamic,rehman2017modified}.
D-QEO bridges this by integrating a true QPU probe as a hardware-level preconditioner.

The strategic gap identified across all these methods is the lack of a high-dimensional (50-qubit) continuous optimization framework that effectively combines quantum global exploration with classical high-precision refinement while managing hardware constraints through the 2-level decomposition. This gap directly motivates the formulation of our approach.

\section{Problem Formulation and Motivation}
\subsection{Constructing the Discretized Hamiltonian: The General Case}
We translate the continuous classical problem into a discrete quantum space through \textbf{binary encoding}.
%leveraging the properties of the Pauli operators. 
A step-by-step mapping for the general case, using the non-separable Himmelblau function~\cite{himmelblau1982reduction} as an  example: 

\subsection*{Step 1: Classical Discretization}
First, we must define the classical boundaries of our search space. Let us define a continuous variable $x$ bounded by a minimum and maximum value: $x \in [x_{min}, x_{max}]$.

If we dedicate $N$ qubits to represent this variable, we can encode $2^N$ discrete states. We can represent any classical integer $k$ in the range $[0, 2^N - 1]$ using standard binary notation:
\begin{equation}
\label{eq:discretization}
    k = \sum_{i=0}^{N-1} q_i 2^i, \quad q_i \in \{0, 1\}
\end{equation}

To map this integer $k$ to our continuous range, we use a simple linear interpolation formula with step size $\Delta = \frac{x_{max} - x_{min}}{2^N - 1}$:
\begin{equation}
x_k = x_{min} + k \Delta
\end{equation}

If we substitute our binary sum into this equation, we get the exact classical value for any bitstring:
\begin{equation}
\label{eq:mapping_val}
x(q) = x_{min} + \Delta \sum_{i=0}^{N-1} q_i 2^i
\end{equation}

\subsection*{Step 2: Formulating the binary number operator}
Next, we formulate an operator whose eigenvalues are classical bits ($q_i \in {0,1}$). 
In quantum mechanics, the computational basis states $|0\rangle$ and $|1\rangle$ are the eigenstates of the Pauli-$\hat \sigma^z$ operator. 
The action of the $\hat{\sigma}^z$ operator on these states yields their corresponding eigenvalues:
\begin{align*}
    \hat \sigma^z|0\rangle &= |0\rangle \quad (\text{eigenvalue is } +1) \\
    \hat \sigma^z|1\rangle &= -|1\rangle \quad (\text{eigenvalue is } -1)
\end{align*}
%\js{you updated Z to $\hat \sigma^z$ below but forgot to do so above} % that was a good catch!!
We require a mathematical transformation that maps the eigenvalues of the Pauli-$\hat \sigma^z$ operator, $\{1, -1\}$, to the classical bit values $q_i \in \{0, 1\}$. 
We achieve this by defining the number operator, $\hat{n}_i$, with the following linear transformation:
\begin{equation}
\hat{n}_i = \frac{\mathbb{I} - \hat{\sigma}^z_i}{2}
\end{equation}
Where $\mathbb{I}$ is the identity matrix, and $\hat{\sigma}^z_i$ is the Pauli-$Z$ operator applied to the $i$-th qubit.

We can verify the eigenvalues of this operator:
\begin{itemize}
    \item If the qubit is in state $|0\rangle$, the $\hat{\sigma}^z$ eigenvalue is $1$. The formula yields: $(1 - 1) / 2 = 0$.
    \item If the qubit is in state $|1\rangle$, the $\hat{\sigma}^z$ eigenvalue is $-1$. The formula yields: $(1 - (-1)) / 2 = 1$.
\end{itemize}

We now have a well-defined quantum operator whose computational basis eigenstates perfectly map to the classical binary values.

\subsection*{Step 3: Building the Variable Operator}
We substitute our number operator ($\hat{n}_i$) back into Eq.~\ref{eq:mapping_val}. 
This gives us a new \textbf{Quantum Operator}, $\hat{X}$, that represents our continuous variable in the Hilbert space:
\begin{equation}
\hat{X} = x_{min}\mathbb{I} + \Delta \sum_{i=0}^{N-1} 2^i \hat{n}_i
\end{equation}

Note that our explicit use of the $\hat{\sigma}^z$ notation alongside operator hats to prevent any notational collision with the Pauli-$X$ operator, $\hat{\sigma}^x$. 
This operator $\hat{X}$ is a matrix whose eigenvalues correspond exactly to the discretized coordinates on our search grid and whose eigenvectors are the computational basis states (the bitstrings).

\subsection*{Step 4: Constructing the Final Hamiltonian}
With the quantum representation for individual continuous variables established, we now substitute these operators into our target objective function to generate the final energy landscape.
The Himmelblau function relies on two variables:
\begin{equation}
    f(x, y) = (x^2 + y - 11)^2 + (x + y^2 - 7)^2
\end{equation}

To evaluate this on a quantum computer, we assign $N$ qubits to represent $x$, and a separate set of $N$ qubits to represent $y$. 
We construct the operator $\hat{X}$ for the first set of qubits, and the operator $\hat{Y}$ for the second set.

Finally, we substitute these operators directly into the Himmelblau polynomial to create our Hamiltonian, $\hat{H}$:
\begin{equation}
    \hat{H} = (\hat{X}^2 + \hat{Y} - 11\mathbb{I})^2 + (\hat{X} + \hat{Y}^2 - 7\mathbb{I})^2
\end{equation}

Expanding this polynomial mathematically yields a sum of tensor products of $Z$ and $\mathbb{I}$ operators across all qubits. 
Because the Hamiltonian is constructed from $\hat{\sigma}^z$ and identity operators, it is not only composed of commuting terms but is also explicitly diagonal in the computational basis. 
Therefore, its gound state is guaranteed to be a single basis state, the bitstring representing the global minimum on our discretized search space. 

% its not only because the hamiltonian 
% its also that the operator is diagonal in the computatational bases
% 

\subsection{The Scaling Challenge of Non-separable Functions}

We evaluated our hybrid algorithm using an asymmetrical Himmelblau function, which features four mathematically identical global minima ($E=0$) at different $x$ and $y$ locations. 

While the classical algorithm distributed its solutions relatively evenly across all four target basins, identifying Global Min 1 $204$ times, Min 2 $280$ times, Min 3 $241$ times, and Min 4 $175$ times, the 5 qubits per dimension hybrid algorithm exhibited severe discretization-induced bias, collapsing all $900$ of its successful runs into Global Min 2.

This behavior is not an algorithmic bug but a fundamental artifact of mapping a continuous search space onto a finite-dimensional Hilbert space. 
By encoding the $[-50,50]$ continuous space using 5-qubits register per dimension, the system is limited to $2^5=32$ orthogonal basis states. 
This discretization forces the quantum state vector to sample the objective function on a coarse grid with $\approx3.2$-unit intervals. 
Due to the alignment of this specific grid, the discrete coordinates nearest to Global Minima 1,3, and 4 land on the steep walls of those valleys, evaluating to high energy floors ($\approx 50.9$, $\approx 14.6$, and $\approx 58.6$, respectively). 

\begin{figure}
    \centering
    \includegraphics[width=0.95\linewidth]{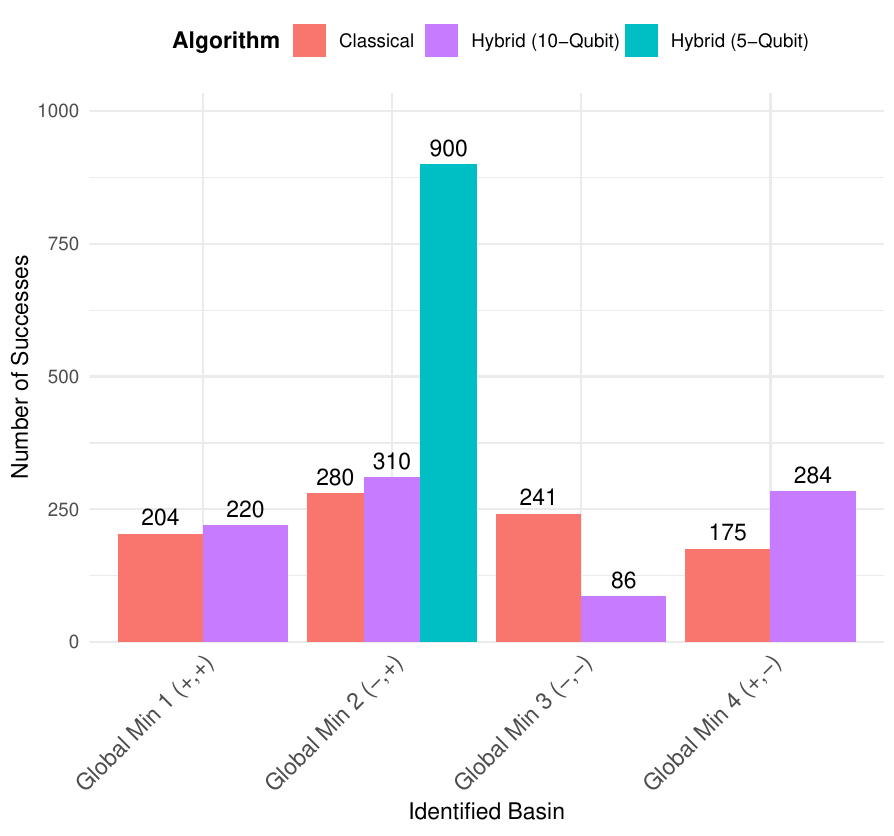}
    \caption{Motivation for cutting and knitting. Increasing from 5 to 10 qubits per dimension removes the grid-alignment bias and restores mathematical symmetry but incurs an exponential scaling cost in qubit requirements.}
    \label{fig:cutting_knitting}
\end{figure}

\begin{figure*}
    \centering
    \includegraphics[width=\textwidth]{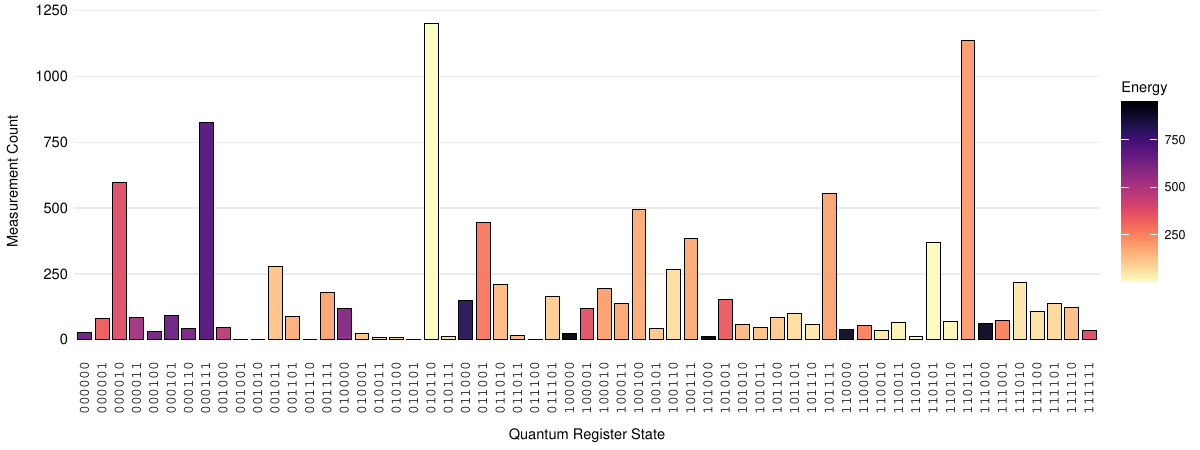}
    \caption{Quantum measurement distribution for the 2D Himmelblau function at low spatial resolution ($K=3$ qubits per dimension). Although the landscape features four global minima ($E=0$), the coarse $8 \times 8$ grid artificially breaks this energy degeneracy. The grid intersection closest to one specific minimum possesses a lower discrete energy than the intersections near the other three. The quantum optimizer correctly identifies this discrete anomaly, resulting in a pronounced wave function collapse into a single basin. Higher qubit resolutions are required to mitigate this grid-induced degeneracy breaking.}
    %probability heatmap mapping a 6-qubit state space to the Himmelblau optimization landscape using 10,000 shots. }
    \label{fig:bitstring}
\end{figure*}

Consequently, the quantum optimizer misses three of the four mathematically identical global minima in the continuous space. 
The coarse discretization artificially breaks this equivalence, skewing the energy landscape and creating a single, dominant global minimum at Global Min 2 ($\approx 6.9$). 
Visually demonstrated in Figure~\ref{fig:bitstring}, this wave function collapses into a single discrete basin (bitstring 010110) as opposed to four corresponding global minima.
To counter this discretization artifact, we increase the register size to 10 qubits per dimension ($1024$ states), drastically refining the grid resolution. 
As shown in Figure~\ref{fig:cutting_knitting}, providing the quantum phase with enough precision and thus distributing the runs much more naturally: $220$ in Min 1, $310$ in Min 2, $86$ in Min 3, and $284$ in Min 4, we were able to find all four global minimum.

However, in general increasing the qubit register is not a scalable solution.  For high-dimensional problems with certain level of entanglement, increasing the qubit count per dimension scales the state space exponentially, quickly exceeding the coherence and connectivity limits of near-term NISQ hardware as well as the memory limits of classical simulators.  This scaling bottleneck directly motivates our future work in quantum circuit cutting and classical tensor network knitting. 
%\ds{By dynamically fragmenting the ansatz into smaller subcircuits (when possible and without paying an exponential overhead for knitting), we hypothesize we can achieve the high-resolution discretization required to accurately model complex, asymmetrical landscapes without violating strict hardware constraints.}

\label{sec:himmelblau}

This exponential scaling bottleneck directly motivates our work in this paper. 
To achieve utility-scale quantum optimization (e.g., 50 qubits) on near-term infrastructure without the cost of classical tensor network knitting, we restrict our focus to \textit{separable} landscapes, leveraging their mathematical structure to perfectly decouple the quantum registers.

\section{Methodology}
The Distributed Quantum-Enhanced Optimization (D-QEO) framework introduces an architectural role reversal in hybrid computing~\cite{fang2010review,sun2012quantum}.
Rather than relying on the quantum processor to pinpoint the exact global minimum, we deploy QPU as a topographical pre-processor. 
Its role is to construct a probability density map of the objective function's landscape, identifying the global basin of attraction. 
These identified regions serve as high-quality seed points for classical high-resolution solvers.

In this work, we restrict our scope strictly to separable functions—encompassing both differentiable (e.g., the Rastrigin function) and non-differentiable (e.g., the Ackley function) landscapes. 
Exploiting this separability enables us to scale the problem through circuit cutting (parallelization) while completely bypassing the exponential overhead associated with classical tensor knitting. 
Despite this structural constraint, our preliminary data (Fig.~\ref{fig:rastrigin_ncorrect}) demonstrates that this approach enhances existing classical PSO methods to a degree that, to the best of our knowledge, remains unachievable without the integration of QPUs.

\subsection{Register-based Mapping}
Unlike particle-based mappings common in classical metaheuristics, D-QEO maps the continuous space into a discrete Hilbert space using a register-based dimensional encoding. 
For a $D$-dimensional optimization problem, we assign a  register of $K$ qubits to each dimension $x_i$. 
Building upon the quantum discretization formulated in Eq.~\ref{eq:mapping_val}, we define the mapping for each dimension $x_i$ using its respective $K$-qubit register:

\begin{equation}
\label{eq:mapping}
\hat{X}_{i} = x_{min}\mathbb{I} + \Delta \sum_{k=0}^{K-1} 2^{k} \hat{n}_{i,k}
\end{equation}

where $\hat{n}_{i,k}$ represents the number operator for the $k$-th qubit in the $i$-th register. For our benchmarks, we utilize $K=5$, allowing a relatively small sub-circuit to explore $2^5 = 32$ discrete sub-regions per dimension.

\subsection{Quantum Topographical Preconditioning with CVaR}
The core of our approach is the Quantum Topographical Preconditioner outlined in Algorithm~\ref{alg:dqeo}. 
To train the variational quantum circuit, we utilize the Conditional Value-at-Risk (CVaR) as our objective function \cite{barkoutsos2020improving}.
For a set of $M=1000$ samples with energies $\{E_1, \dots E_M\}$ sorted in ascending order, we define the objective for a confidence level $\alpha=0.1$ \cite{barkoutsos2020improving} as:

\begin{equation}
\text{CVaR}_{\alpha}(\theta) = \frac{1}{\lceil \alpha M \rceil} \sum_{k=1}^{\lceil \alpha M \rceil}E_k
\end{equation}

The classical optimizer iteratively updates the variational parameters to minimize this tail energy.
As the optimization progresses, the quantum wave-function naturally ``bunches'' around the lowest energy eigenstates, which correspond to the most promising basins of attraction in the classical landscape.

%\ds{As visually demonstrated in Figure~\ref{fig:bitstring} this preconditioning effect successfully maps the 2D Himmelblau landscape without suffering from severe mode collapse. Across $10,000$ shots the $8 \times 8$ quantum measurement grid reveals distinct probability spikes strictly isolated to the four structurally identical global minima. Because these continuous minima reside at non-integer coordinates, the rigid $K=3$ discrete grid misses the absolute bottom of each basin by varying spatial distances. This creates a \textit{discretization bias}, artificially lowering the discrete evaluation of one basin relative to the others.}

\begin{figure}[htbp]
\small
\centering
\begin{tikzcd}[row sep=0.5cm, column sep=0.2cm]
\lstick{$q_0$} & \gate{H} & \gate{R_y(\theta_0)} & \ctrl{1} & \qw      & \qw      & \qw      & \targ{}  & \gate{R_y(\theta_5)} & \ctrl{1} & \qw \dots & \meter{} \\
\lstick{$q_1$} & \gate{H} & \gate{R_y(\theta_1)} & \targ{}  & \ctrl{1} & \qw      & \qw      & \qw      & \gate{R_y(\theta_6)} & \targ{}  & \qw \dots & \meter{} \\
\lstick{$q_2$} & \gate{H} & \gate{R_y(\theta_2)} & \qw      & \targ{}  & \ctrl{1} & \qw      & \qw      & \gate{R_y(\theta_7)} & \qw      & \qw \dots & \meter{} \\
\lstick{$q_3$} & \gate{H} & \gate{R_y(\theta_3)} & \qw      & \qw      & \targ{}  & \ctrl{1} & \qw      & \gate{R_y(\theta_8)} & \qw      & \qw \dots & \meter{} \\
\lstick{$q_4$} & \gate{H} & \gate{R_y(\theta_4)} & \qw      & \qw      & \qw      & \targ{}  & \ctrl{-4}& \gate{R_y(\theta_9)} & \qw      & \qw \dots & \meter{} 
\end{tikzcd}
\caption{Hardware-Efficient Ansatz utilized for a single $K=5$ qubit dimension register. The circuit builds a parameterized probability distribution over the discrete search space.}
\label{fig:ansatz}
\end{figure}
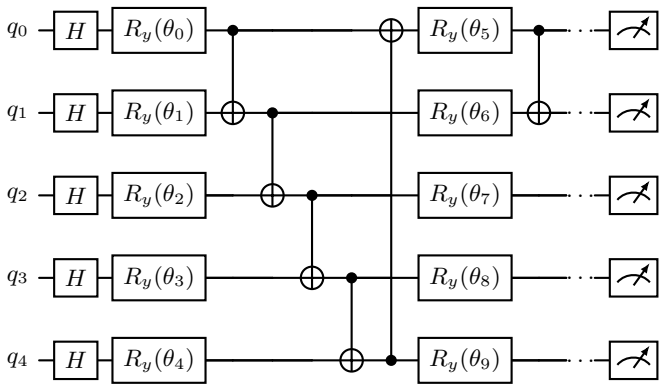

% cite the top papers that use this ansatz
% why we used this? its the sota
% 
To prepare this state, we employ a Hardware-Efficient Ansatz (HEA) \cite{barkoutsos2020improving,stenger2025hybrid} (and references within) that has an initial layer of Hadamard gates to create a uniform superposition, followed by  parameterized $R_y$ rotation layers ($l=3$) and a cyclic CNOT entangling ring. The ansatz for a single dimension and $K=5$ qubit register is illustrated in Figure~\ref{fig:ansatz}.

\subsection{Distributed Circuit Execution for Separable Landscapes}
A defining feature of this methodology is the exploitation of functional separability.
For a separable objective function $f(x) = \sum_{i=1}^{D} f_i(x_i)$, the corresponding global Hamiltonian decomposes perfectly into $D$ independent partial Hamiltonians ($\hat H_i$) and thus the total is: $\hat{H}_{total} = \sum_{i=1}^{D} \hat{H}_{i}$.

Consequently, the global quantum state does not require cross-register entanglement. 
We can mathematically and operationally cut the $D \times K$ qubit circuit into $D$ independent $K$-qubit fragments.
This guarantees the distribution and the evaluation of each dimensional fragment asynchronously across multiple QPUs + GPUs (or just GPUs in our CUDA-Q simulation) with $O(c)$ quantum knitting overhead, essentially bypassing the exponential complexity of high-dimensional quantum optimization.
To simulate utility-scale environments (e.g., 50 qubits) in our experiments, we employ this mathematical decomposition alongside circuit-level cutting using CUDA-Q~\cite{kim2023cuda}.
Because the quantum fragments are perfectly decoupled, the memory requirements of each independent $K$-qubit subcircuit fit within the VRAM of a single GPU.
This allows execution of the global evaluation by concurrently processing the subcircuit fragments on a single GPU.
By doing so, we maximize classical simulation throughput and demonstrate that high-dimensional quantum optimization can be achieved without requiring massive supercomputing. 
%Of course, this has future implications when hardware QPUs and GPUs will be available to our project.  
%\ds{This decoupled architecture is natively compatible with future QPU-GPU hardware deployments, allowing dynamic workload distribution across distinct physical processors.}

\subsection{Hybrid VQE and Classical Refinement}
The transition from the quantum environment to the classical HPC solver requires translating the localized quantum wave-functions back into continuous floating-point variables. 
Since we cannot pass the raw measurement bitstrings directly to the classical optimizer, we extract the statistical properties of the optimized probability distribution to define a highly bounded, continuous search space.

Once the variational training is complete, we perform a global concatenation to reconstruct the seed coordinates for classical refinement. We extract the absolute best coordinates found $x_{best}$, and calculate the weighted centroid of the samples in the CVaR tail, $x_{cvar}$. The final seed coordinate is computed as a weighted average:

\begin{equation}
x_{seed} = \beta \times\mathbf{x}_{best} + (1-\beta) \times x_{cvar}
\end{equation}

Inspired by PSO~\cite{kennedy1995particle}, $\beta$ serves as a hyperparameter, controlling the trust placed in the single best measurement versus the distributional centroid.
Through experimentation, we picked $\beta=0.7$. Future thorough investigation will be performed. 

To construct the bounding box for the classical solver, we define a dynamic search radius $\delta$. 
Drawing on the principles of step-size adaptation in Covariance Matrix Adaptation (CMA-ES)~\cite{hansen2016cma} and stochastic trust-region methods~\cite{conn2000trust}, $\delta$ scales with the root-mean-square (RMS) spatial deviation:

\begin{equation}
\delta = \delta_{base} + \gamma \times RMS
\end{equation}

We set the minimum algorithmic trust-region $\delta_{base} = 0.5$ to match the $\pm 0.5$ boundaries of our target global basins ($\S$~\ref{sec:evaluation_metrics}). By setting the scaling multiplier $\gamma = 2$, the expansion term acts as a $2\sigma$ spatial confidence interval. This dynamically expands the classical solver's bounds strictly when the quantum sub-routines exhibit high spatial uncertainty. 
These bounded coordinates initialize the classical phase~\cite{soos2025zeus}.

\begin{algorithm}[ht]
\caption{D-QEO: Quantum Topographical Preconditioner}
\label{alg:dqeo}
\begin{algorithmic}
\Require Dimensions $D$, Qubits per dimension $K$, Bounds $[X_{min},X_{max}]$, separable objective $f(x) = \sum f_i(x_i)$
\State \textbf{Phase 1: Initialization \& Distribution}
\State Initialize $D$ independent quantum registers of size $K$
\State Map each $f_i(x_i)$ to local Hamiltonian $\hat{H}_i$ using Eq. (\ref{eq:mapping})
\State \textbf{Phase 2: Parallel Quantum Preconditioning}
\For{$i = 1$ to $D$} \textbf{in parallel} 
    \State Initialize Ansatz $\lvert \Psi_i(\theta_i) \rangle$ (Fig. \ref{fig:ansatz})
    \While{convergence criteria not met}
        \State Sample 1000 shots from $\lvert \Psi_i(\theta_i) \rangle$ once per iteration
        \State Calculate $\text{CVaR}_{0.1}$ from  lowest 100 energy samples
        \State Update $\theta_i$ using COBYLA to minimize CVaR
    \EndWhile
    \State Extract $x_{best,i}$ and calculate tail centroid $x_{cvar,i}$
\EndFor
\State \textbf{Phase 3: Reconstruction \& Classical Refinement}
\State Construct $X_{seeds}$ and calculate search radius $\delta$ with variance
\State Decode $B$ into continuous coordinates $X_{seeds}$
\If{$f(x)$ is differentiable}
    \State Warm-start PSO+BFGS at $X_{seeds}$ within radius $\delta$
\Else
    \State Warm-start PSO around $X_{seeds}$ within radius $\delta$
\EndIf
\State \Return Global Minimum found by classical solver~\cite{soos2025zeus}
\end{algorithmic}
\end{algorithm}

\section{Experiments}

To evaluate the performance of the proposed Hybrid D-QEO architecture, we benchmark our solver against classical baselines~\cite{soos2025zeus} using a set of highly non-convex, symmetric optimization functions. 
These landscapes are chosen for the dense population of local minima, which are designed to trap standard optimization algorithms. 

\subsection{Experimental Setup and Hardware Configuration}
To ensure the reproducibility of our results, all quantum circuit simulations were executed using NVIDIA's CUDA-Q framework. 
The simulations were performed in state-vector mode, accelerated by a single NVIDIA A100 Tensor Core GPU (80 GB). 
This hardware configuration allowed for the efficient classical simulation of our 50-qubit space by executing the independent 5-qubit chunks.

\subsubsection{Evaluation Metrics and Statistical Trials} \label{sec:evaluation_metrics}
Because stochastic optimization algorithms exhibit variance, each experiment was repeated for 100 independent trials. 
The primary metric for success, $N_{correct}$, is defined as the number of trials that successfully converge to the true global minimum. 
A trial is classified as ``correct'' if the final optimized coordinates fall strictly within the basin of attraction of the global minimum (e.g., bounded within $\pm 0.5$ of the origin across all dimensions\cite{soos2025zeus}).
%This ensures that the algorithm has fundamentally bypassed the surrounding local minima rather than simply finding a local trap.

\subsubsection{Search Space Reduction Metrics}
To rigorously quantify the topographical advantage of the quantum preconditioner, we track the continuous bounds generated by the quantum state. 
Across the 100 independent trials for a given dimension $D$, we extract the widest (worst-case) bounding box $[lb, ub]$ that successfully captured the global basin. 
From this, we calculate the preconditioned search volume as $V_{pre} = \prod_{i=1}^{D} (ub_i - lb_i)$.

Furthermore, because landscapes like Rastrigin and Ackley are highly symmetric, with local minima occurring at  approximate integer intervals, we calculate the total number of local minima within these reduced bounds. 
For a given dimension $i$, the number of trapped minima is determined by the number of integers bounded within $[lb_i, ub_i]$. 
The total preconditioned minima count is the product of these surviving traps across all $D$ dimensions (see Table~\ref{tab:volume_reduction}). 

\subsection{Differentiable landscapes}
For the differentiable continuous benchmark, we test the algorithm on the $N$-dimensional Rastrigin function.
Rastrigin is difficult for classical solvers due to its cosine-modulated landscape, which produces thousands of deep local minima surrounding a single global minimum.

We demonstrate that utilizing the quantum phase as a preconditioner fundamentally alters the search dynamics. 
By collapsing the probability wave into the bounding box containing the true global minimum, the quantum phase bypasses the high-frequency local minima. 
As we increase the number of dimensions, the classical baseline severely drop in the number of correct solutions. 
In contrast, the hybrid algorithm maintains a higher percentage of correct solutions. 

To analyze the impact of the quantum optimizer's computational budget on the final convergence variance, we employ the gradient-free COBYLA~\cite{powell1994direct} optimizer to train the variational circuit during Phase 1. 
It is well-documented in optimization literature that the performance of gradient-free algorithms like COBYLA is heavily dependent on the available iteration budget, often requiring a substantial number of functional evaluations to yield high-quality results \cite{bagheri2017self, powell1994direct, rios2013derivative}. 
To quantify this trade-off between the depth of the quantum search and the stability of the final classical convergence, we systematically limit the quantum preconditioning phase to maximum functional budgets of $N_{eval} \in \{200, 2000, 8000\}$.

\subsection{Non-differentiable landscapes}
In this setting, we explore the algorithmic trade-off of optimizing highly non-convex landscapes where the classical gradient is undefined. 
Purely classical optimization on such functions requires gradient-free approximation or purely heuristic methods, which suffer from severe computational overhead and slow convergence rates in high dimensions. 

By utilizing our quantum preconditioner, which evaluates the landscape with discrete state sampling rather than relying on local gradients, we can effectively isolate the target basin without requiring objective function differentiability. 
We also demonstrate that feeding this tightly constrained bounding box to a classical gradient-free solver reduces the search volume exponentially.

The following section presents the empirical outcomes of these trials, specifically focusing on dimensional scaling, computational effort, and topological volume reduction.

\section{Results}
\subsection{Convergence Stability Across Dimensions}
\begin{figure}[ht]
    \centering
    \includegraphics[width=\linewidth]{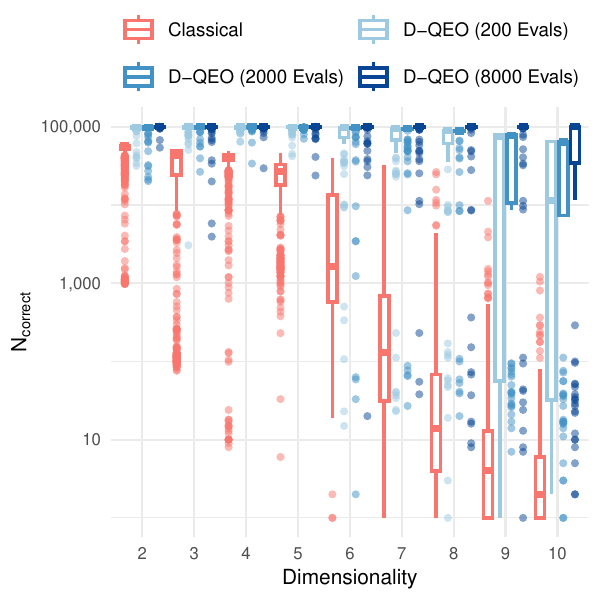}
    \caption{Number of correct solutions ($N_{correct}$) for the Rastrigin function. The D-QEO preconditioner preserves convergence in high dimensions where the classical baseline experiences exponential failure. The box plots display the distribution of successful runs. The central line indicates the median, the box edges represent the 25th and 75th percentiles (Interquartile Range), the whiskers extend to $1.5 \times \text{IQR}$, and the discrete dots represent outlier runs.}
    \label{fig:rastrigin_ncorrect}
\end{figure}

We first evaluate the algorithm's ability to locate the true global minimum basin as the dimensionality of the problem scales. 
Figure~\ref{fig:rastrigin_ncorrect} illustrates the number of successful solutions ($N_{correct}$) achieved by the classical baseline compared to the D-QEO on the Rastrigin function. 
As is consistent with established literature regarding the curse of dimensionality \cite{soos2025zeus} (and references within), the purely classical algorithm exhibits an exponential decay in success rate, failing almost entirely as the search space reaches $10$ dimensions.
Conversely, the quantum preconditioner drastically stabilizes the swarm. 
By collapsing the probability wave into the bounding box containing the true global minimum, the hybrid algorithm maintains a remarkably high percentage of correct solutions regardless of the ambient dimensional scaling.

Even at the restricted quantum evaluation budget of $N_{eval} = 200$, the hybrid algorithm maintains a reliable success rate in 10 dimensions. 
Increasing the variational budget to $N_{eval} = 8000$ further tightens the probability distribution of the quantum state, resulting in a near-perfect success rate across all evaluated dimensions. 
This behavior directly aligns with established classical findings that the solution quality of gradient-free algorithms like COBYLA scales heavily with the available evaluation budget \cite{bagheri2017self, powell1994direct, rios2013derivative}. 
However, our results demonstrate a key quantum advantage: even an ``early,'' truncated COBYLA run provides sufficient topographical data to the quantum state to effectively warm-start and rescue the classical solver.

\begin{figure}[ht]
    \centering
    \includegraphics[width=\linewidth]{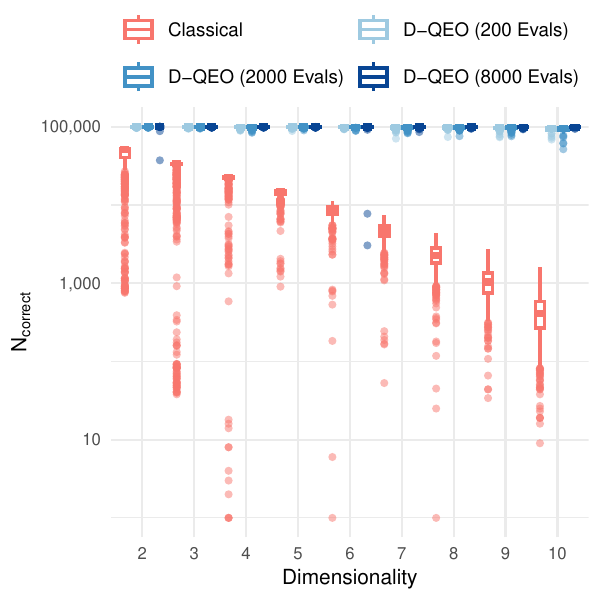}
    \caption{Number of correct solutions ($N_{correct}$) for the separable Ackley function. The quantum sampling effectively bypasses the classical gradient failures caused by the function's non-differentiability at the global optimum. Plotting conventions  follow those defined in Figure \ref{fig:rastrigin_ncorrect}.}
    \label{fig:ackley_ncorrect}
\end{figure}

To demonstrate the versatility of this preconditioning approach, we also evaluated the algorithms on the separable Ackley function shown in Figure~\ref{fig:ackley_ncorrect}.
The Ackley landscape is highly difficult for quasi-Newton methods because its non-differentiability at the origin.
Because the quantum preconditioner evaluates the landscape with discrete state sampling rather than local gradients, it effectively bypasses this non-differentiability. 
As a result, the D-QEO warm-started classical solver consistently achieves a higher yield of correct solutions in high dimensions compared to the purely classical baseline.

\subsection{Reduction in Classical Computational Effort}
\begin{figure}[ht]
    \centering
    \includegraphics[width=\linewidth]{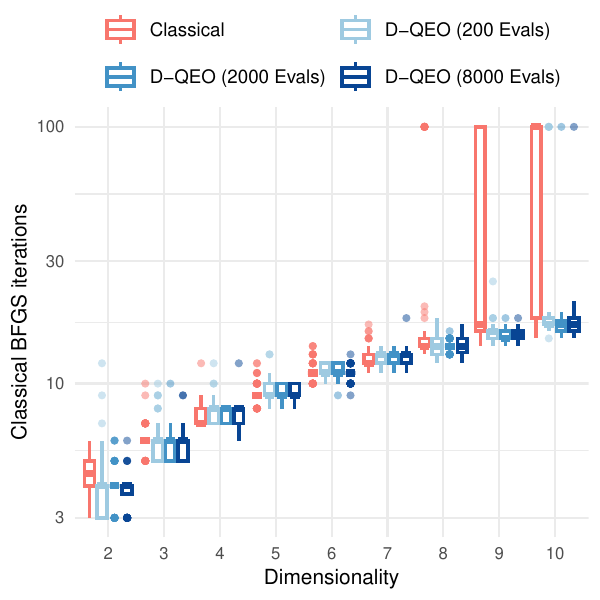}
    \caption{Classical optimization effort required to achieve convergence. The quantum topographical warm-start drastically reduces the number of classical BFGS iterations needed to traverse the landscape, indicating a shift from dissipative classical computation to efficient quantum preconditioning. Plotting conventions follow those defined in Figure \ref{fig:rastrigin_ncorrect}.}
    \label{fig:bfgs_iterations}
\end{figure}

In addition to improving accuracy, D-QEO must demonstrate a tangible reduction in the classical workload. 
Figure~\ref{fig:bfgs_iterations} tracks the classical computational effort—measured in total BFGS iterations—required to achieve convergence.

It is important to note that because these experiments are executed on classical GPU-based quantum simulators, we do not claim a reduction in total wall-clock execution time, as simulating 50 qubits inherently introduces massive classical overhead.
Rather, these results demonstrate a fundamental reduction in \textit{algorithmic complexity} and classical pathlength.
By utilizing the QPU to map the topography and identify the global basin of attraction, the D-QEO framework effectively skips the classical exploration phase. 

Consequently, the classical BFGS optimizer requires fewer iterations to descend to the exact minimum compared to an un-preconditioned swarm.
This validates the conjecture that increasing qubit resolution replaces dissipative, energy-intensive classical iterations with energy-preserving quantum unitary operations, thereby fundamentally reducing the energy footprint required for complex global optimization.

\subsection{Volume Reduction}
Beyond classical iteration counts, the mathematical advantage of the D-QEO preconditioner is best understood through the reduction of the search space and the systematic elimination of local minima. 
Table~\ref{tab:volume_reduction} compares the original classical search space against the worst-case localized bounding boxes extracted from the quantum distribution. 

The results demonstrate staggering efficiency gains that fundamentally disrupt the ``curse of dimensionality''. 
For instance, in the 10-dimensional Rastrigin landscape, the ambient space spans $[-5.12, 5.12]^D$, containing $11^{10} \approx 2.59 \times 10^{10}$ (over 25 billion) local minima. 
However, the quantum preconditioner bounds the search so tightly around the global basin that in the worst-case trial, 7 of the 10 dimensions were restricted to a width of less than 1.0, trapping only the true minimum coordinate ($0$). 
The remaining 3 dimensions exhibited slightly more variance, trapping two minima each. 

Consequently, the quantum warm-start collapsed a 25-billion minima problem into a localized continuous subspace containing only $1^7 \times 2^3 = 8$ total local minima. 
By feeding this tightly constrained bounding box to the classical solver, the quantum phase mathematically eliminates the vast majority of the landscape's non-convexity, trapping the classical optimizer within a continuous subspace where failure is highly improbable. 
Similar exponential reductions are observed in the non-differentiable Ackley landscape, validating the framework's robustness across varied topographies.

\begin{table*}[htbp]
\centering
\caption{Topographical Preconditioning Impact: Comparing the original classical search space against the constrained bounding boxes ($[\mathbf{lb}, \mathbf{ub}]$) extracted from the quantum distribution. The reduction factor illustrates the exponential advantage of isolating the global basin.}
\label{tab:volume_reduction}
\begin{tabular}{lc|cc|c|cc}
\toprule
\textbf{Landscape} & $D$ & \textbf{Orig. Vol. ($V_{orig}$)} & \textbf{Precond. Vol. ($V_{pre}$)} & \textbf{Reduction Factor} & \textbf{Orig. Minima} & \textbf{Precond. Minima} \\
\midrule
Rastrigin & 2 & $104.86$ & $2.80$ & \textbf{$37.46$} & $121$ & $2$ \\
 & 3 & $1,073.74$ & $6.90$ & \textbf{$155.53$} & $1,331$ & $8$ \\
 & 4 & $1.10 \times 10^{4}$ & $11.08$ & \textbf{$992.10$} & $1.46 \times 10^{4}$ & $4$ \\
 & 5 & $1.13 \times 10^{5}$ & $19.09$ & \textbf{$5,896.43$} & $1.61 \times 10^{5}$ & $8$ \\
 & 6 & $1.15 \times 10^{6}$ & $41.75$ & \textbf{$2.76 \times 10^{4}$} & $1.77 \times 10^{6}$ & $8$ \\
 & 7 & $1.18 \times 10^{7}$ & $108.64$ & $\bf 1.09 \times 10^{5}$ & $1.95 \times 10^{7}$ & $64$ \\
 & 8 & $1.21 \times 10^{8}$ & $76.60$ & $\bf1.58 \times 10^{6}$ & $2.14 \times 10^{8}$ & $8$ \\
 & 9 & $1.24 \times 10^{9}$ & $228.67$ & $\bf5.41 \times 10^{6}$ & $2.36 \times 10^{9}$ & $16$ \\
 & 10 & $1.27 \times 10^{10}$ & $178.41$ & $\bf 7.11 \times 10^{7}$ & $2.59 \times 10^{10}$ & $8$ \\
\midrule
Ackley & 2 & $4,294.97$ & $28.15$ & \textbf{$152.57$} & $4,225$ & $30$ \\
 & 3 & $2.81 \times 10^{5}$ & $162.13$ & \textbf{$1,736.10$} & $2.75 \times 10^{5}$ & $150$ \\
 & 4 & $1.84 \times 10^{7}$ & $3,398.70$ & \textbf{$5,427.60$} & $1.79 \times 10^{7}$ & $3,584$ \\
 & 5 & $1.21 \times 10^{9}$ & $5,394.95$ & \textbf{$2.24 \times 10^{5}$} & $1.16 \times 10^{9}$ & $5,400$ \\
 & 6 & $7.92 \times 10^{10}$ & $2.44 \times 10^{4}$ & \textbf{$3.24 \times 10^{6}$} & $7.54 \times 10^{10}$ & $1.88 \times 10^{4}$ \\
 & 7 & $5.19 \times 10^{12}$ & $1.39 \times 10^{5}$ & $\bf 3.75 \times 10^{7}$ & $4.90 \times 10^{12}$ & $1.62 \times 10^{5}$ \\
 & 8 & $3.40 \times 10^{14}$ & $5.73 \times 10^{5}$ & $\bf 5.94 \times 10^{8}$ & $3.19 \times 10^{14}$ & $9.33 \times 10^{5}$ \\
 & 9 & $2.23 \times 10^{16}$ & $4.73 \times 10^{6}$ & $\bf 4.71 \times 10^{9}$ & $2.07 \times 10^{16}$ & $4.86 \times 10^{6}$ \\
 & 10 & $1.46 \times 10^{18}$ & $3.63 \times 10^{7}$ & $\bf 4.03 \times 10^{10}$ & $1.35 \times 10^{18}$ & $4.03 \times 10^{7}$ \\
\bottomrule
\end{tabular}
\end{table*}

\section{Discussion and Future Work}

\subsection{Algorithmic Complexity vs. Hardware Overhead}
The results demonstrate a clear reduction in the classical computational effort required to solve high-dimensional separable landscapes. 
However, it is critical to contextualize these findings within the current state of quantum computing. 
Currently, the total wall-clock execution time of the D-QEO framework is dominated by the overhead of classically simulating utility-scale quantum circuits. 

Therefore, this study explicitly evaluates \textit{algorithmic complexity} and classical iteration reduction rather than raw temporal speedup. 
The framework is designed with the foresight that as physical QPUs mature—specifically regarding gate execution times, active qubit reset capabilities, and QPU-CPU network bandwidth—the physical wall-clock execution will naturally align with the theoretical algorithmic efficiency demonstrated in this study.

\subsection{NISQ Constraints and Hardware Noise}
It is also worth noting that the primary scaling analysis presented in this work is using CUDA-Q. 
%% is an idealized algorithmic study assuming perfect statevector evolution. 
By the time of the presentation, we anticipate that we will have data from a subset of IBM Nighthawk or Heron processors.
%In near-term implementations, gate infidelities, thermal relaxation, and readout errors will dampen the precision of the variational probability mapping. 

Because D-QEO utilizes the QPU strictly as a coarse-grained topographical preconditioner rather than a high-precision solver, we hypothesize it inherently possesses a high tolerance for such errors. 
%However, a comprehensive analysis evaluating the exact impact of realistic hardware noise profiles on the CVaR objective landscape remains a critical next step for validation.

\subsection{Extending to Non-Separable Landscapes}
While this work successfully circumvents the exponential spatial bottleneck by restricting its scope to separable functions, many high-value industrial~\cite{abi2020volume}
and basic research %~\cite{xyz} 
optimization problems are inherently non-separable, featuring heavily coupled variables (as demonstrated by the Himmelblau scaling failure in Section~\ref{sec:himmelblau}). 

Future work will focus on extending the D-QEO architecture to navigate these highly coupled landscapes. 
To achieve this, we plan to explore advanced dynamic circuit cutting and classical tensor network knitting techniques. 
By strategically severing the cross-register entanglement operations, we aim to reconstruct the high-resolution probability amplitudes to break the symmetry of complex, non-separable functions without exceeding the strict coherence and connectivity limits of NISQ-era architectures. 
Furthermore, future iterations will prioritize the integration of Quantum Error Mitigation (QEM) protocols to facilitate direct execution on physical quantum hardware.

\section{Conclusion}
In this paper, we presented the Distributed Quantum-Enhanced Optimization framework, a novel hybrid architecture designed to mitigate the ``curse of dimensionality'' in continuous global search. 
By strategically reversing the traditional roles of hybrid computing, we deployed the QPU as a topographical preconditioner to construct probability density maps of the objective landscape, leaving high-precision continuous refinement to classical GPU solvers. 
To achieve utility-scale execution, we exploited the mathematical structure of separable continuous functions, allowing a monolithic 50-qubit search space to be mathematically severed and distributed into independent subcircuits without incurring exponential tensor-knitting overhead. 

Our experiments on the highly non-convex Rastrigin and non-differentiable Ackley functions demonstrate that D-QEO successfully neutralizes the exponential failure rates inherent to purely classical swarms. 
By collapsing the quantum wave function into the bounding box of the true global minimum, the quantum warm-start dramatically reduces the required classical pathlength and BFGS iteration count. 
Ultimately, this framework provides a highly scalable, distributed blueprint for translating near-term quantum resources into tangible computational reductions for high-dimensional optimization.

\section*{Acknowledgment}
This research was supported in part by the Fermi National Accelerator Laboratory (FERMILAB-CONF-26-0212-CSAID) graduate Summer Internship Program (DS \& MP), the Office of Naval Research (ONR) through the U.S. Naval Research Laboratory (JS), and the Richard T. Cheng Endowment (NC) at ODU. 
The authors thank Min Dong at ITS in ODU for his help with CUDA-Q at ODU. 
The authors would like to acknowledge the use of Google's Gemini AI during the preparation of this manuscript. 
In accordance with IEEE policy, we note that this generative AI system was utilized strictly as an advanced copyediting and formatting assistant. 
All scientific concepts, experimental data, algorithms, and core analytical arguments remain the entirely original, human-generated work of the authors. 
This work was performed using computational facilities at ODU enabled by grants from the National Science Foundation (MRI grant no. CNS-1828593) and Virginia's Commonwealth Technology Research Fund and Google Cloud Platform through ODU's Monarch Sphere initiative. 
%This work was produced by FermiForward Discovery Group, LLC under FERMILAB-CONF-26-0212-CSAID.
Any subjective views or opinions expressed in this paper do not necessarily represent the views of the national labs, the NSF, or the United States Government.

\bibliographystyle{IEEEtran}
\bibliography{references}

\end{document}